\documentclass[journal]{IEEEtran}
\usepackage{cite}

\ifCLASSINFOpdf
\else
  \usepackage[dvips]{graphicx}
  \DeclareGraphicsExtensions{.eps}
\fi
\usepackage[cmex10]{amsmath}
\usepackage[tight,footnotesize]{subfigure}

\usepackage{fixltx2e}

\usepackage{stfloats}
\begin{document}
%
\title{Ultra stable and very low noise signal source using a cryocooled sapphire oscillator for VLBI}
%
%
%

\author{Nitin~R.~Nand, John~G.~Hartnett, Eugene~N.~Ivanov and Giorgio ~Santarelli,
\thanks{John~G.~Hartnett, ~Nitin~R.~Nand and ~Eugene~N.~Ivanov are with the School of Physics, the University of Western Australia, Crawley, 6009, W.A., Australia.  ~Giorgio ~Santarelli is with LNE-SYRTE, Observatoire de Paris, CNRS, UPMC, 61 Avenue de l$\,\acute{}$Observatoire, 75014 Paris, France}
\thanks{Manuscript received March 23, 2011; ....}}

%
%

\markboth{IEEE Trans. on Microwave Theory and Techniques,~Vol.~XX, No.~X, December~2011}%
{Shell \MakeLowercase{\textit{et al.}}: Bare Demo of IEEEtran.cls for Journals}
%



\maketitle

\begin{abstract}
Here we present the design and implementation of a novel frequency synthesizer based on low phase noise digital dividers and a direct digital synthesizer. The synthesis produces two low noise  accurate and tunable signals at 10 MHz and 100 MHz.  We report on the measured residual phase noise and frequency stability of the synthesizer, and estimate the total frequency stability, which can be expected from the synthesizer seeded with a signal near 11.2 GHz from an ultra-stable cryocooled sapphire oscillator. 

The synthesizer residual single sideband phase noise, at 1 Hz offset, on 10 MHz and 100 MHz signals, respectively, were measured to be -135 dBc/Hz and -130 dBc/Hz. Their intrinsic frequency stability contributions, on the 10 MHz and 100 MHz signals, respectively, were measured as $\sigma_y = 9 \times 10^{-15}$ and $\sigma_y = 2.2 \times 10^{-15}$, at 1 s integration time. 

The Allan Deviation of the total fractional frequency noise on the 10 MHz and 100 MHz signals derived from the synthesizer with the cryocooled sapphire oscillator, may be estimated as 
$\sigma_y \approx 5.2\times10^{-15} \tau^{-1}+3.6\times10^{-15}\tau^{-1/2}+ 4\times 10^{-16}$
and
$\sigma_y \approx 2\times 10^{-15} \tau^{-1/2}+3\times 10^{-16}$, respectively, for 1 s $\leq \tau < 10^4$ s. 

We also calculate the coherence function, (a figure of merit in VLBI) for observation frequencies of 100 GHz, 230 GHz and 345 GHz, when using the cryocooled sapphire oscillator and an hydrogen maser. The results show that the cryocooled sapphire oscillator offers a significant advantage at frequencies above 100 GHz.

\end{abstract}


%
\IEEEpeerreviewmaketitle

\section{Introduction}
\IEEEPARstart{V}{ery-long-baseline-interferometry} (VLBI) radio astronomy requires a low-phase-noise ultra-stable frequency reference. Such a reference is commonly  derived at 10 MHz  from a hydrogen maser.  The development of a cryocooled sapphire oscillator (cryoCSO) and frequency synthesizer described here offers a lower noise alternative. 

Cryogenic sapphire oscillators (CSO) \cite{chang,hart1,locke, hartnitin1,hartnitin2, grop1, grop2} have demonstrated extremely good frequency stability (characterized in terms of Allan deviation) at integration times in the range 1 s to 1000 s. However, these oscillators do not operate at precisely prescribed repeatable frequencies. Transferring their stability to  precisely 10 MHz and 100 MHz presents technical challenges \cite{chambon1,chambon2,boudot, doeleman}. 

In this paper, we describe the design and implementation of a synthesizer based on an ultra-stable CSO \cite{wang,hartnitin1,hartnitin2} complemented by  very low phase noise digital dividers and a direct digital synthesizer (DDS) \cite{DDS}.  We evaluate the performance of the synthesizer in terms of phase noise and stability of the synthesized 10 MHz and 100 MHz reference signals.   
 
\begin{figure}[!t]
\centering
\includegraphics[width=3.5in]{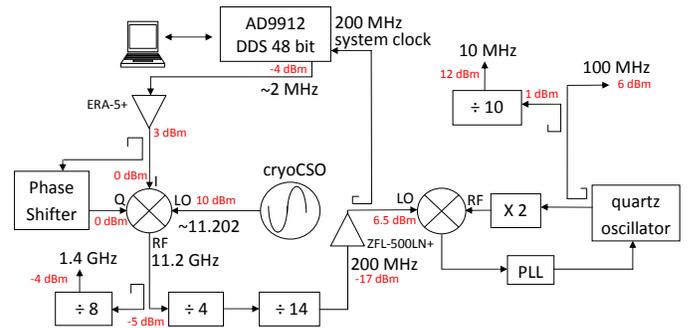}
\caption{A block diagram of the frequency synthesizer. Attenuators, filters, and DC blocks not shown. The principle components used were the Analog Devices AD9912 DDS module, the Marki IQ-0714 mixer, the low phase noise Wenzel 501-04517D quartz oscillator, the low phase noise Holzworth HX4210 divider ($\div 10$), the Minicircuits amplifiers ERA-5+ and ZFL-500LN+, the Minicircuits frequency multiplier ZX90-2-13-S+, the Hittite dividers  HMC-C007 ($\div$ 8), HMC365S8G ($\div$ 4) and the programmable HMC705LP4 divider (where we used $\div$ 14). Note: PLL (Phase-Locked Loop).}
\label{fig1}
\end{figure}
  
\section{Frequency synthesizer design and measurement techniques}  
\subsection{Synthesizer design and architecture} 
We constructed two nominally identical frequency synthesizers. Fig. 1 shows a block diagram of the synthesizer based on a stable microwave oscillator, the cryoCSO, and a high-resolution DDS phase referenced to a sub-harmonic of the cryoCSO signal. The output signal frequency of the cryoCSO is $11.201 \, 967\, 189\, 42$ GHz. This signal is not tunable and the closest integer multiple of 100 MHz is 11.2 GHz. By  frequency shifting the cryoCSO signal with a $1.967$ XXXXX MHz signal, generated by the DDS unit, this produces the correct frequency for integer digital frequency division.
The frequency shifting was performed with an image rejection mixer in order to suppress the spurious mixing product about 4 MHz above the useful signal. Without this the spurious signal would impair the divider operation and filtering it out would require the use of an impractically high-Q-factor microwave filter. The use of the image rejection mixer is an elegant alternative to solve the problem. 

By combining an X-band IQ commercial mixer with a custom made $\pi$/2 phase shifter we realized an  image rejection of about 40 dB. The signal levels are optimized to suppress the carrier by about 40 dB. At the output of the image rejection mixer a 2-step  digital frequency division by 56 produces a 200 MHz output. This 200 MHz  signal, down-converted   from the cryoCSO, is used to clock the DDS synthesizer, as well as a reference signal for the phase-locked loop controlling the frequency of a low phase noise 100 MHz quartz oscillator, which is frequency doubled by a commercial low phase noise frequency multiplier. The phase-locked loop ensures that the ultra-high frequency stability of the cryoCSO is transferred to the 100 MHz quartz oscillator with only a small addition of noise. The frequency of the latter is further divided to 10 MHz, but with a non-negligible noise contribution from the divider.

Furthermore the circuit provides a 1.4 GHz by dividing the output signal of the mixer  by 8. This signal, which is very close to the hydrogen  maser microwave transition, is a good candidate for future very low noise synthesizers. It is worth noting that the technique described here can easily be adapted to any of the possible microwave output  frequencies from the cryogenic sapphire oscillator.

\begin{table} 
\begin{center}
	\caption{SSB phase noise (${\mathcal L}_{\varphi}$) @ 1 Hz offset for the active components used in the frequency synthesizer at the power levels and frequencies$^{(vi)}$ specified in Fig. 1.}
		\begin{tabular}{c c c c}
Component 	 					  						&${\mathcal L}_{\varphi}$   &Output freq.  &${\mathcal L}_{\varphi}$ @\\
Model Number												&[dBc/Hz]		&	MHz				&100 MHz$^{(v)}$\\
				\hline 	
				\hline																																				
HMC-C007 ($\div$ 8)										&-120 &1,400  	&  \\		
HMC365S8G ($\div$ 4)$^{(i)}$					&     &2,800 	&  \\
HMC705LP4	($\div$ 14)		 							&-124 &200		& -130\\
HX4210($\div$ 10)											&-135 &10  		&  \\
ERA-5+	amp														&-120 &2		 	&-161\\
ZFL-500LN+ amp$^{(ii)}$								&$<$ -144 &200  	&$<$ -150 \\	
AD9912 DDS														&-103 &2 			&-144\\	
501-04517D	quartz$^{(iii)}$ 					&-70 &100     		&-134\\	
cryoCSO$^{(iv)}$											&-97 &11,202  &-138\\
\hline
		\end{tabular}
	\label{actdevpn}
\end{center}
$^{(i)}$ Not measured individually, only used as input to HMC705LP4.\\
$^{(ii)}$ Only individually measured at 100 MHz, where its phase noise could not be measured above the measurement noise floor.\\
$^{(iii)}$ The free running phase noise for the oscillator at 100 Hz offset is -130 dBc/Hz \cite{sprinter} and with a $f^{-3}$ dependence ${\mathcal L}_{\varphi}$ = -70 dBc/Hz at 1 Hz. In the last column the residual phase noise inside the PLL is shown where it was measured using two oscillators locked to a common 100 MHz signal with a locking bandwidth of about 100 Hz. \\
$^{(iv)}$ Absolute phase noise from \cite{hartnitin2}\\
$^{(v)}$ We are interested in their contribution on the 100 MHz carrier at the output of the synthesizer.\\
$^{(vi)}$ For the ERA-5+ amp, the DDS and the cryoCSO division by 56 times means 35 dB reduction to their contribution at 200 MHz and 41 dB at 100 MHz.\\
\end{table}

\subsection{Measurement techniques}
Phase noise measurements are traditionally made using the analog technique by mixing equal frequency signals 90$^{\circ}$ out of phase as shown in Fig. 2(a). The  baseband signal voltage is then sampled and Fourier analyzed by a FFT spectrum analyzer, and then post-processed to extract the phase noise data. 

Initially, we used the analog technique to evaluate the residual phase noise of the components used to realize the synthesizer. See Fig. 3. In a second phase we used a Symmetricom 5125A phase noise test set; hereafter referred to as the test set (see Fig. 2). The test set outputs the single sideband (SSB) phase noise and the signal frequency stability of the device under test (DUT) in real-time by comparing it with a reference signal. The test set performs phase detection by digital signal processing (DSP) methods, sampling the RF waveforms directly \cite{testset}. 

\begin{figure}[!t]
\centering
\includegraphics[width=3.5in]{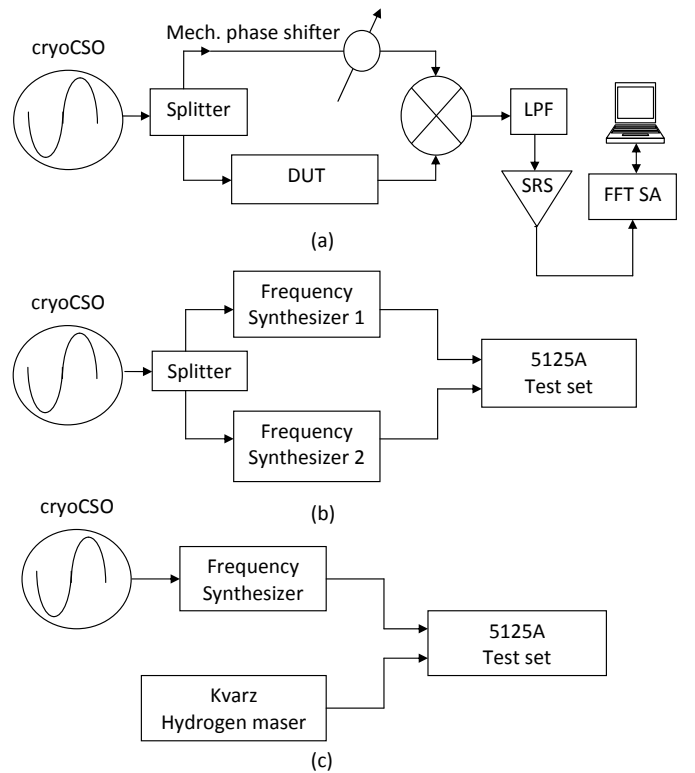}
\caption{Set-up for measuring; (a) residual phase noise of the Device Under Test (DUT) via the  analog technique, (b) residual phase noise and stability (Allan deviation) via a Symmetricom 5125A phase noise test set, and (c) stability between a  hydrogen maser and synthesizer via a Symmetricom 5125A phase noise test set. Note: LPF (Low Pass Filter), SRS (Stanford Research Systems Model SR560 low noise preamplifier), and FFT SA (Agilent 89410A Fast Fourier Transform Spectrum Analyzer).}
\label{fig2}
\end{figure}

\section{Results and discussion}
Oscillator phase noise can be described by a power law given by
\begin{equation}
S_{\varphi}(f)=\sum^{-4}_{i=0}b_{i}f^{i},
\label{pnlaw}
\end{equation}
where $S_{\varphi}(f)$ is the power spectral density of phase fluctuations $\varphi (t)$ and has units rad$^{2}$/Hz. In the literature, phase noise is commonly reported as
\begin{equation}
{\mathcal L}_{\varphi}(f)=\frac{1}{2} S_{\varphi}(f)= 10\,Log_{10} [S_{\varphi}(f)]-3\,dB,
\label{lpn}
\end{equation}
which has units dBc/Hz. In this paper all phase noise measurements are reported in ${\mathcal L}_{\varphi}(f)$ or SSB units. On a log-log plot, the term $f^{i}$ maps on to a straight line of slope i $\times$ 10 dB/decade. The value of the slope is used to identify the various noise processes commonly encountered in oscillators \cite{rubi}.


\subsection{Residual phase noise and stability}
\subsubsection{Residual phase noise}
We measured the residual phase noise of the DDS, dividers, and amplifiers (as the DUT) using set-up Fig. 2(a) and show the residual phase noise of the DDS and dividers in Fig. 3. Table \ref{actdevpn} summarizes the relevant phase noise contributions of the active components used in the design of the synthesizer.  The power spectral density of phase fluctuations on the DDS 2 MHz output signal  is -103 dBc/Hz at 1 Hz offset. Since these fluctuations are superimposed on the microwave carrier (when DDS signal is mixed with that of the cryoCSO) their power spectral density is reduced to -144 dBc/Hz when microwave frequency is divided to 100 MHz and is negligible in the total noise budget of the synthesizer. Its contribution is less than the phase noise of the cryoCSO signal referred to 100 MHz, which is -138 dBc/Hz at 1 Hz offset \cite{hartnitin2}.  The last column of Table \ref{actdevpn} lists the relevant noise contributions of each component referred to 100 MHz.

Using the cryoCSO output signal to drive the inputs of two nominally identical synthesizers we separately measured the relative phase noise (see Fig 2(b)) of the 10 MHz and 100 MHz output signals. The results are shown in Fig. 4 in curves (1) and (2), respectively, and specify the residual phase noise of a single synthesizer. Curve (3) is the phase noise of a single 100 MHz-10 MHz digital frequency divider (model Holzworth HX4210, curve (4) from Fig. 3) used here. Curves (4) and (5) are the measurement noise floors for the test set at 10 MHz and 100 MHz, respectively. Curve (6) indicates the expected level of phase noise from the cryoCSO signal when divided down to 100 MHz. It contributes negligibly to the phase noise of the synthesizer at 100 MHz. The indicated spurs at multiples of the 1.46 Hz of the cryocooler compressor cycle result from poor rejection by the measurement equipment. This was proven when the cryocooler compressor, used to cool the resonator in our loop oscillator, was switched off yet another was still running nearby and the same spurs were observed. Note the power levels of these spurs are about 20 dB lower on the 10 MHz signal than on the 100 MHz signal. The spur near 60 kHz is the residual Pound modulation sideband from the cryoCSO loop oscillator.

\begin{figure}[!t]
\centering
\includegraphics[width=3.5in]{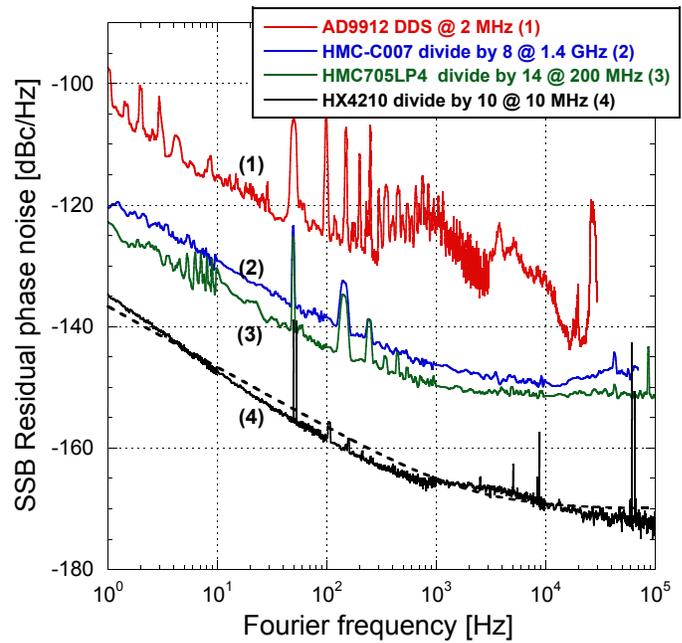}
\caption{(Color online) Phase noise of some of the components used in our design measured via the analog technique (Fig 2(a)). Curve (1) the AD9912 DDS, curve (2) the HMC-C007 divide by 8, curve (3) the HMC705LP4 divide by 14 where the input was derived from a HMC365S8G divide by 4. Curve (4) is the phase noise of the HX4210 divider, with 1 dBm input power, measured using the 5125A test set and the set-up of Fig. 2(a) with the divider as the DUT. The broken line is the measured phase noise for the HX4210 divider taken from the manufacturer's data sheet. }
\label{fig3}
\end{figure}

\begin{figure}[!t]
\centering
\includegraphics[width=3.5in]{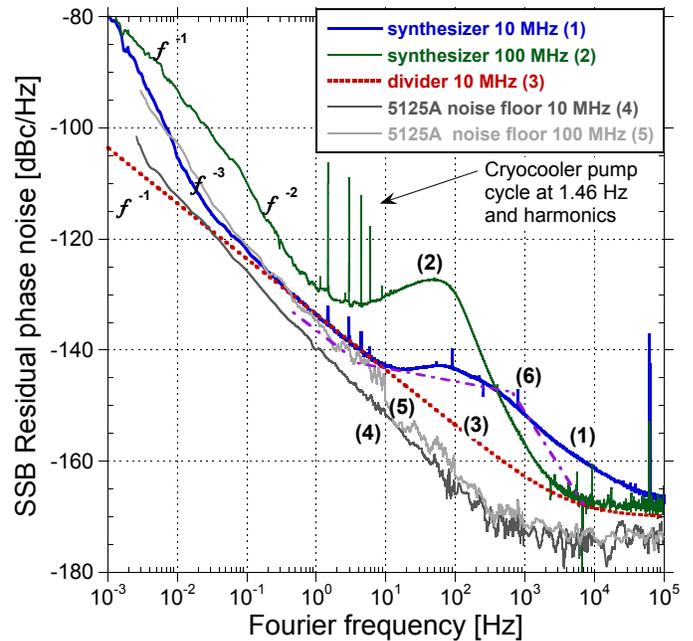}
\caption{(Color online) The single sideband (SSB) residual phase noise of the synthesized 10 MHz (curve (1)) and 100 MHz (curve (2)) reference signals measured with a Symmetricom 5125A phase noise test set.  Data shown represent a single synthesizer, from the relative noise from two nominally identical units. Curve (3) (dotted line) is the phase noise of a single 100 MHz - 10 MHz divider and curves (4) and (5) are the measurement noise floors in the test set at 10 MHz and 100 MHz respectively. Curve (6) (the dot dashed line) is an estimate of the level of phase noise of the cryoCSO signal divided down to 100 MHz.}
\label{fig4}
\end{figure}

From Fig. 4, one can clearly see that the bandwidth of the PLL controlling the 100 MHz quartz oscillator is about 100 Hz. At Fourier frequencies $f >100$ Hz the phase noise of the 100 MHz output  signal (curve (2)) is due to the free-running quartz oscillator (which is consistent with its specifications). At $f<3$ Hz, i.e. well within the PLL bandwidth, the phase noise of the 100 MHz output signal is due to intrinsic fluctuations of the RF mixer used in the PLL. In addition to the PLL mixer, the low frequency noise is also due to the intrinsic fluctuations in the frequency divider (division by 56) and in the 200 MHz amplifier. These are sources of uncorrelated phase fluctuations in two synthesizers. See Table \ref{actdevpn} for their contributions on the 100 MHz carrier.


The residual phase noise of the 10 MHz signal (curve (1)) is clearly limited by the frequency divider phase noise at Fourier frequencies  $2 \times 10^{-2}$ Hz $ < f < 10$ Hz. The 10 MHz synthesizer phase noise  is flicker phase dominated, and at Fourier frequencies $f< 2 \times 10^{-2}$ Hz it exhibits flicker frequency noise. For $f < 100$ Hz the shape of the  the phase noise spectrum of the 10 MHz signal can be explained by the residual phase noise of the 100 MHz output reduced by 20 dB (the relative frequency reduction) plus the intrinsic noise of the divider, but for $f>300$ Hz it is not clear why we see some excess noise in the 100 MHz-10 MHz frequency divider. We are currently investigating the effect of changing the operating temperature setpoint of the dividers and whether we can explain this.

From these data the SSB residual phase noise at 1 Hz offset on the 10 MHz and 100 MHz signals  is -135 dBc/Hz and -130 dBc/Hz, respectively. The former is confirmed by the phase noise measurement of a single divider as indicated by (4) in Fig. 3. A very good 10 MHz quartz oscillator phase noise is -122 dBc/Hz at 1 Hz offset \cite{quartz10}. Our result here is 13 dB lower than the best quartz.

\subsubsection{Frequency stability}

Using the set-up of Fig. 2(b), we measured the intrinsic frequency stability of the synthesized 10 MHz and 100 MHz references (curves (1) and (2) of Fig. 5). The measurements obtained with the test set that are shown in Fig. 5 are for a Nyquist equivalent noise bandwidth (NEQ BW) of 0.5 Hz.  The NEQ BW determines the minimum integration or averaging time ($\tau_{0}$) for stability measurements and is given by $\tau_{0}=1/(2\,\text{NEQ\,BW})$. (See Ref. \cite{testset} for details.) The test set tabulates the Allan deviation of the fractional frequency fluctuations at certain integer multiples of $\tau_0$. These data are used in the figures herein. 

Curves (3) and (4) are the measurement noise floors of the test set, measured by splitting the outputs of the synthesizer with a power divider. These were then subtracted from curves (2) and (3) and the intrinsic stability ($\sigma_y(\tau)$) contribution for a single  synthesizer at 10 MHz and 100 MHz calculated.  At 1 s of integration, these are $9 \times 10^{-15}$ and $2.2 \times 10^{-15}$,  for the 10 MHz and 100 MHz outputs, respectively. The degradation in stability on the 10 MHz signal can be attributed to the phase noise contribution of the 100 MHz-10 MHz divider itself (see Fig. 4). If the division to 10 MHz was noiseless we would expect the same fractional frequency stability on that signal as on the input 100 MHz signal.

It should be noted that the 100 MHz to 10 MHz frequency divider is  temperature sensitive and the intrinsic stability shown in Fig. 5 is the best result obtained while actively controlling the temperature of the divider with a thermistor and resistive patch heater a few degrees above the ambient temperature.  
 
Using the set-up of Fig. 2(c) we measured the stability of the cryoCSO/synthesizer against our hydrogen maser at 100 MHz. The resulting Allan deviation of the combined fractional frequency fluctuations is given by curve (1) in Fig. 6. It is the best result we measured over a two week period.  The hydrogen maser noise dominates the measurements out to about $10^4$ s of averaging. The circled data indicate the cryoCSO frequency stability where it is assumed that the cryoCSO dominates over that of the maser. To fully characterize the frequency instability of the synthesized signals for $\tau < 10^4$ s a second cryocooled oscillator is needed.

From the previously measured frequency stability of the cryoCSO against a liquid helium cooled CSO \cite{hartnitin2} (curve (2) in Fig. 6) it is possible to estimate the expected short term and very long term stability of a single cryoCSO (curve (3) in Fig. 6). This estimate is based on the assumption that for times $\tau < 100$ s both oscillators are equal in performance and at times $\tau \geq 10^5$ s the frequency stability of the cryoCSO is that of curve (1), from the comparison of the cryoCSO with our maser. This must be the case else we would see the long term stability (no frequency drift removed) similar to that of curve (2) for $\tau > 10^3$ s. For $\tau > 100$ s the liquid helium cooled CSO frequency stability is worse than that of the cryoCSO.  Hence we are able to calculate the total noise contribution from the cryoCSO and a single synthesizer for integration times $\tau < 100$ s, where we have reliable data. 

In order to calculate the expected stability of the synthesized signals with the cryoCSO the frequency stability of the cryoCSO (curve (3) of Fig. 6) was added to the intrinsic stability of the synthesizer (curves (1) and (2) of Fig. 5) at 10 MHz and 100 MHz, respectively.  The results are shown in Fig. 7 where they are compared with the frequency stability of our maser (curve (1) for $\tau < 10^5$ s) and with that of a high performance maser (curve (4)).  Curves (2) and (3) are the frequency stabilities for the 10 MHz and 100 MHz cryoCSO/synthesizers, respectively.

This calculated data was interpolated to the measured comparison data between the hydrogen maser and the cryoCSO/synthesizer (curve (1) in both Figs 6 and 7) for $\tau \geq 10^{5}$ where the noise contribution is assumed to be largely due to the cryoCSO. As a result we were able to curve fit to the data of curves (2) and (3), in Fig. 7, resulting in flicker floors estimated to be about $4 \times 10^{-16}$ (10 MHz) and $3 \times 10^{-16}$ (100 MHz) at integration times around 1000 s. These flicker floors cannot be well specified due to insufficient data.

\begin{figure}[!t]
\centering
\includegraphics[width=3.5in]{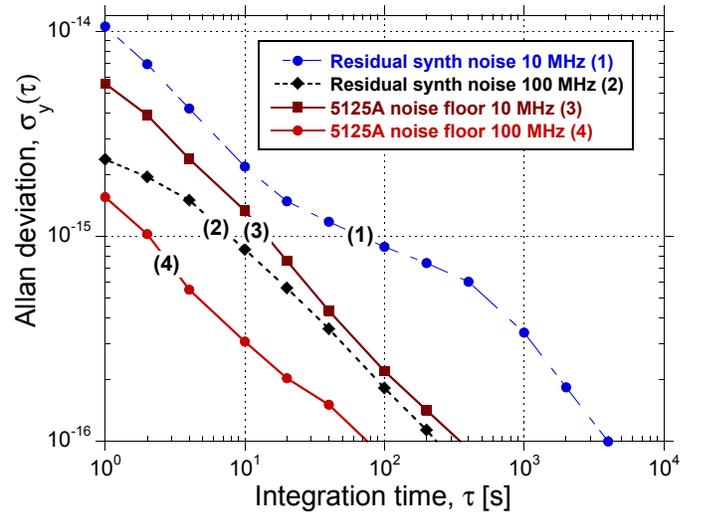}
\caption{(Color online) The Allan deviation of the measured frequency stabilities for a Nyquist equivalent noise bandwidth of 0.5 Hz: curves (1) and (2) represents  the intrinsic stability of the 10 MHz and the 100 MHz synthesizers, and curves (3) and (4) are the measurement noise floors at 10 MHz and 100 MHz respectively.}
\label{fig5}
\end{figure}

\begin{figure}[!t]
\centering
\includegraphics[width=3.5in]{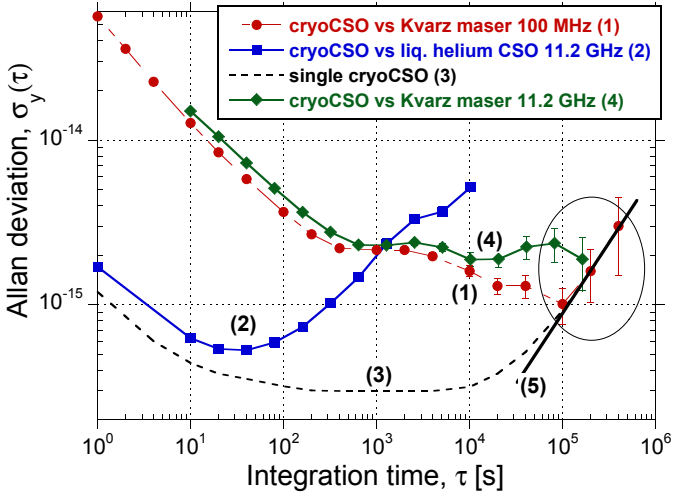}
\caption{(Color online) The Allan deviation of the measured frequency stabilities. Curve (1) represents the 100 MHz synthesizer directly compared with with the 100 MHz output our Kvarz hydrogen maser and curve (2)  the cryoCSO compared with a liquid helium cooled CSO at 11.2 GHz. Curve (3) is the expected stability of a single cryocooled CSO. Curve (4) represents the Allan deviation calculated for the times series data of Fig. 8.  The solid line labeled (5) represents the long term $\tau$ dependence from which a maximum value of frequency drift is calculated.}
\label{fig6}
\end{figure}

\begin{figure}[!t]
\centering
\includegraphics[width=3.5in]{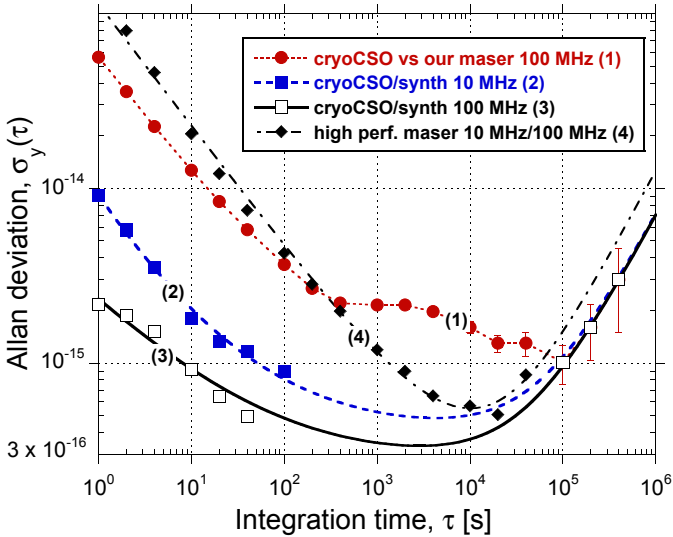}
\caption{(Color online) Best measured stability between our Kvarz hydrogen maser and the cryoCSO/synthesizer at 100 MHz (solid circles, curve (1)). The curves (2) and (3) are estimates due to the cryoCSO/synthesizer alone, at 10 MHz and 100 MHz. respectively. Curve (4) is the typical stability data of a very high performance hydrogen maser. }
\label{fig7}
\end{figure} 

However, since we only have one cryocooled sapphire oscillator we were not able to directly measure its stability for $60$ s $ < \tau < 10^4$ s. A second cryocooled oscillator is necessary to do this. The fits (curves (2) and (3) in Fig. 7) represent the expected, yet optimistic, total stability of the synthesizer using the cryoCSO and may be described by

\begin{eqnarray}
&\sigma_{y}(\tau)_{10} = 5.2\times10^{-15} \tau^{-1}+3.6\times10^{-15}\tau^{-1/2}+ ... \nonumber \\
& + 4\times 10^{-16} +7\times 10^{-21} \tau, \label{fittotal10}
\end{eqnarray}
and
\begin{eqnarray}
&\sigma_{y}(\tau)_{100}=2.1\times 10^{-15} \tau^{-1/2}+3\times 10^{-16}+ ... \nonumber \\
& +7\times 10^{-21} \tau, \label{fittotal00} 
\end{eqnarray}
where the subscripts represent the particular output frequency of the synthesizer in MHz. These fits are valid for $\tau > 10^5$ s.

The long term frequency stability at both 10 MHz and 100 MHz signals should converge due to the long term stability of the cryoCSO. However beyond $10^4$ s the cryoCSO stability is largely determined by the environment; temperature and pressure changes. Also there is an uncertainty about the frequency stability of the hydrogen maser, so the only way to truly establish this performance is with at least one more cryocooled sapphire oscillator.

\subsection{Long term performance} 
The cryoCSO signal frequency is subject to drift due to ambient pressure and temperature changes in the lab, though the lab temperature is stable to $\pm 0.1^\circ$ C. These changes translate into  temperature and pressure changes within the cryostat. The design of the cryostat has a mixed liquid-helium-gas space. For a schematic of the cryostat see Fig. 1 of Ref. \cite{hartnitin2}. This means the region where the cryocooler condenser constantly reliquefies a small quantity of helium gas in a closed system. It has been found that best operating condition in this helium gas space is where the pressure is maintained as low as possible. 

Figure 7 (curve (1)) shows a comparison at 100 MHz of the cryoCSO/synthesizer compared to our hydrogen maser. This data was taken over a period of about 2 weeks after the cryoCSO had been continuously operating for about 9 months. It has been observed that the data exhibit a long term trend of decreasing frequency. This measurement represents the quietest data segment in terms of low frequency drift that we have taken to date. The solid line (labeled (5)) in Fig. 6 is a fit used to determine the drift rate of $8\times 10^{-16}$/day assuming drift dominates over random walk of frequency. As the cryoCSO was improved on, a 4 K radiation shield reduced thermal gradients \cite{hartnitin2} and reduced frequency drift but the origin of the remaining frequency drift is unknown.

The normal operating pressure inside the helium gas space is about 46 kPa. A steady state of helium gas being reliquefied maintains this. If power is shut off to the compressor, for example, in the event of a power failure,  the pressure in this region will begin to rise.  This was observed, when, due to a maintenance issue, the power to the whole lab was shut off for about 20 minutes. After  the power was restored the pressure stabilized to its normal condition within an hour and the oscillator started automatically and remained functioning.

Subsequently we measured the time evolution of the cryoCSO frequency after this event. This was measured by down-converting the 11.202 GHz signal of the cryoCSO to approximately 2 MHz with a 11.200 GHz signal produced from a higher order harmonic of a Step-Recovery diode driven with a doubled 100 MHz output of our hydrogen maser. A similar technique was used in Ref. \cite{chambon2}. The 2 MHz output from the mixer was filtered and directly counted with an Agilent 53132A counter with a 10 s gate time. This method was necessary to obtain a time series of the evolution of the beat between the cryoCSO and the microwave reference signal from our maser.  See Fig. 8 for the fractional frequency offset from the moment the cryoCSO oscillator started up again. The drift rate is negative and decreasing. The best fit to 7 days of data is described by,
\begin{equation}
\frac{\Delta f}{f} = a (e^{-t/t_0}-1),
\label{drift}
\end{equation}
where $a = (5.79 \pm 0.02) \times 10^{-14}  $ and $t_0 = 4.44 \pm 0.02$ days. By the 7th day Eq. (\ref{drift}) represents a decreasing fractional frequency drift in the cryoCSO frequency of $2.7 \times 10^{-15}$/day.  Since we observed the decreasing exponential frequency drift for many months from the initial start-up, after cooling from room temperature, it cannot be attributed to a change in pressure in the helium gas space. That quickly stabilizes in a matter of hours. The cause of this long term exponential decrease is not known as yet. 

The frequency stability calculated from the time series data is shown in Fig. 6 (curve (4)) with error bars. The exponential frequency drift (Eq. (\ref{drift})) was removed before the Allan deviation was calculated. The multiplication process via the step-recovery diode used to generate the 11.2 GHz from the 100 MHz maser signal and amplification adds some noise, and also there seems to be an unknown temperature dependence there as well. Nevertheless, the stability calculated from the 7 days of data of Fig. 8 shows very good reproducability of the cryoCSO performance even after a 20 minute power interruption.

\begin{figure}[!t]
\centering
\includegraphics[width=3.5in]{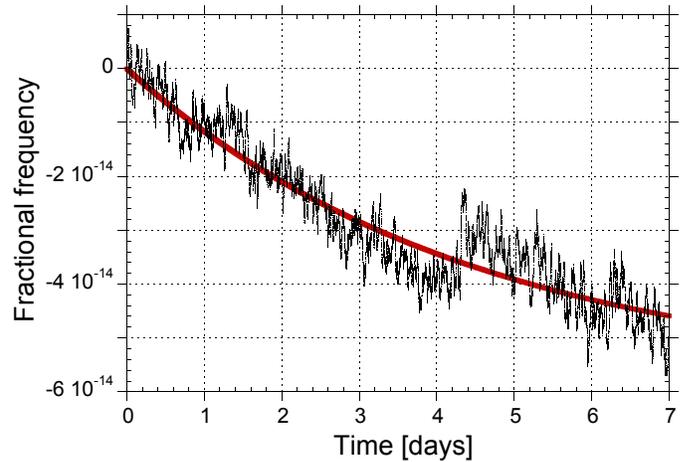}
\caption{(Color online) Evolution of the measured fractional frequency of the cryoCSO compared with that of a hydrogen maser up to 7 days after the power to the lab was interrupted then restored (data are averages over 1000 s). The solid line is the best exponential fit to the data.}
\label{fig8}
\end{figure}

\section{VLBI coherence}
In Very-Long-Baseline-Interferometry radio astronomy, the coherence $C(T)$, a function of integration time $T$, is defined as 
\begin{equation}
C(T) = \left| \frac{1}{T}\int^{T}_{0}  \text{Exp}\left[i \phi(t)    \right] dt \right|,
\label{coh}
\end{equation}
where $\phi(t)$ is the phase difference between the two stations forming the interferometer. Considering only the phase difference due to the stability of the frequency standards used, the value of the coherence function $C(T)$ for an integration time $T$ (henceforth referred to as coherence time) is a good figure of merit and can be estimated from, \cite{VLBIcoherence1,VLBIcoherence2}
\begin{equation}
\left\langle C(T)^{2}\right\rangle=\frac{2}{T}\int^{T}_{0}\left(1-\frac{\tau}{T}\right) \text{Exp}\left[ -\frac{\omega^{2}\,\tau^{2}}{4}\sum^{n}_{i = 0}\sigma^{2}_{y}(2^i \tau)\right] d\,\tau,
\label{rmscoh}
\end{equation}
where $\omega$ is the angular frequency of the local oscillator (equal to the frequency of the astronomical source) and $\sigma^{2}_{y}(\tau)$ is the local reference signal Allan variance at integration time $\tau$. The sum inside the square brackets should converge as $n \rightarrow\infty$, which would be true for frequency standards limited by white phase noise. With the standards used here this is not actually the case and the expression must be cut off at some point. In our case  it was sufficient to use  n = 3 at 345 GHz and  n = 4  at 230 GHz and 100 GHz for calculations with all frequency standards. See Ref. \cite{VLBIcoherence2} for further details.
 
For a particular standard at a given observing frequency, the coherence function has the range 0 to 1. $C(T)$ is proportional to the fringe amplitude \cite{fringe} and when $\left\langle C(T)^{2}\right\rangle^{1/2}$ = 1 there is no loss of coherence. The coherence time is approximately the time for which the coherence function $\left\langle C(T)^{2}\right\rangle^{1/2}$ approaches 1. 

\begin{figure}[!t]
\centering
\includegraphics[width=3.5in]{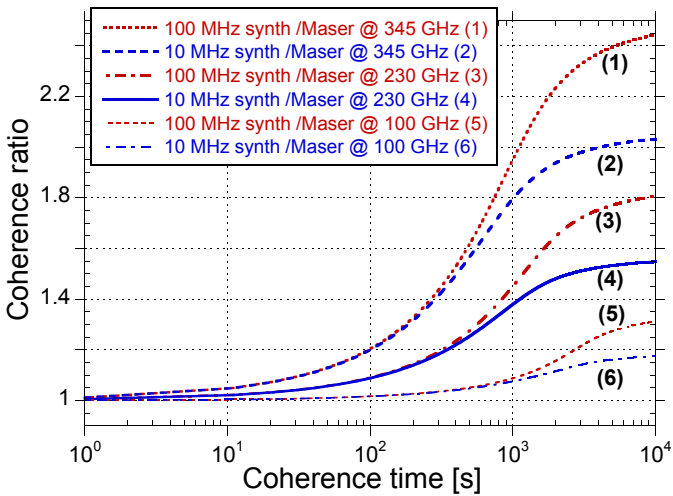}
\caption{(Color online) Ratio of the coherence function calculated for both the 10 MHz and 100 MHz outputs of the cryoCSO/synthesizer with that for the  maser data. The stability of the maser is the same for both the 10 MHz and the 100 MHz outputs.}
\label{fig9}
\end{figure}

\begin{table}
\begin{center}
	\centering
	\caption{Allan deviation of reference signals and calculated Coherence for an observing frequency of 345 GHz with the cryoCSO and a high performance maser }
		\begin{tabular}{c c c c c }
 		&$\sigma_{y}(\tau)$ 	&$\left\langle C(T)^{2}\right\rangle^{1/2}$  		&$\sigma_{y}(\tau)$		&$\left\langle C(T)^{2}\right\rangle^{1/2}$  	 \\
\hline
\hline
$\tau$ 	&cryoCSO  	&cryoCSO			&cryoCSO			&cryoCSO \\
(or T)						&100 MHz 		&100 MHz			&10 MHz &10 MHz \\
				\hline 																																					
$1$ 			&$2.4 \times 10^{-15}$ 	&0.99999		&$9.2 \times 10^{-15}$ 	&0.99991\\		
$10$			&$9.4 \times 10^{-16}$ 	&0.99994 	&$2.1 \times 10^{-15}$ 	&0.99964\\
$10^2$		&$4.9 \times 10^{-16}$ 	&0.99825   &$8.1 \times 10^{-16}$ 	&0.99513\\
$10^3$		&$3.5 \times 10^{-16}$  &0.91664		&$5.3 \times 10^{-16}$ 	&0.84457\\
$10^4$		&$3.7 \times 10^{-16}$	&0.39452		&$5.1 \times 10^{-16}$	&0.32815\\
		\hline 
$\tau$		&Maser 			&Maser		\\
(or T)					&10/100 MHz		&10/100 MHz				\\
				\hline 																																					
$1$ 			&$1.1 \times 10^{-13}$	&0.98862		\\		
$10$			&$2.3 \times 10^{-14}$  &0.95524		\\
$10^2$		&$4.8 \times 10^{-15}$	&0.82920		\\
$10^3$		&$1.2 \times 10^{-15}$	&0.47022		\\
$10^4$		&$5.5 \times 10^{-16}$	&0.16146		\\	
\hline
			\end{tabular}
	\label{coherence}
\end{center}
\end{table}

From a qualitative point of view the coherence time is shorter at higher observing frequencies. Therefore the stability of the local reference will impact more strongly at higher frequencies over the shorter averaging times. Hence for millimeter wave observations the stability of the local oscillator becomes very important assuming the seeing conditions (the atmosphere) is not the limitation.

Using the analytical expressions, Eqs (\ref{fittotal10}) and (\ref{fittotal00}), obtained from fits to our measured stability $\sigma_{y}(\tau)$ data, in Fig. 7, we numerically calculated the coherence function from Eq. (\ref{rmscoh}). Hence $\left\langle C(T)^{2}\right\rangle^{1/2}$ was determined using our 10 MHz and 100 MHz references, multiplied up to millimeter wave frequencies of 100 GHz, 230 GHz and 345 GHz. It is assumed that the same performance local oscillator is used on each end of the interferometer and that the multiplication process does not add any noise. The results are listed for 345 GHz in Table \ref{coherence} along with the oscillator stability at fixed integration times and compared to that calculated from a very high performance  maser. A typical value of the T-4 Science maser \cite{T4maser} has been assumed. The frequency stability for that  maser is shown as curve (4) in Fig. 7.

Similarly the coherence function $\left\langle C(T)^{2}\right\rangle^{1/2}$ was calculated using the stability data for this hydrogen maser  (with the same stability at both 10 MHz and 100 MHz)  where the same assumption was made as above.  The resulting coherence function is then compared by plotting the ratio of $\left\langle C(T)^{2}\right\rangle^{1/2}$ derived from either the 10 MHz or 100 MHz synthesizer with that when the maser is used. The results are shown in Fig. 9, and show that the calculated coherence function is greater when using the 100 MHz reference due to its better short term stability. 

The references synthesized from the cryoCSO are significantly more frequency stable than those from a hydrogen maser, especially the 100 MHz output. The latter offers improvements above 200\% in the value of the coherence function $\left\langle C(T)^{2}\right\rangle^{1/2}$ at observing frequencies of 345 GHz at coherence times near $10^4$ s, assuming that the frequency reference is the limitation. It is apparent from Fig. 9 that at observing frequencies less than 100 GHz there is only a small advantage in using the cryoCSO. It must be said that where the decoherence effects of the atmosphere can be avoided, or compensated for, there is a clear advantage in integrating  the signals for an hour or more. This is where the cryoCSO offers a big advantage over the hydrogen maser, especially for millimeter wave VLBI.

\section{Conclusion}
Two nominally identical frequency synthesizers based on low phase noise digital dividers and a direct digital synthesizer have been constructed and their performance evaluated. The reference signals at 10 MHz and 100 MHz were synthesized from a cryocooled sapphire oscillator and their phase noise and stability measured.  The synthesizer residual single sideband phase noise, at 1 Hz offset, on 10 MHz and 100 MHz signals, respectively, were measured to be -135 dBc/Hz and -130 dBc/Hz. Their intrinsic frequency stability contributions, on the 10 MHz and 100 MHz signals, respectively, were measured as $\sigma_y = 9 \times 10^{-15}$ and $\sigma_y = 2.2 \times 10^{-15}$, at 1 s integration time. As such the fractional  frequency noise on 100 MHz output, at short integration times, is only about a factor of 2 greater than that of the cryocooled oscillator itself. The estimated total frequency stabilities of the new references are significantly better than those for the same output frequencies from a very high performance hydrogen maser. 

From these measurements, we calculated the coherence function, a figure of merit, for millimeter wave VLBI radio astronomy.  The references synthesized from the cryocooled sapphire oscillator offer improvements in terms of the coherence function of the order of 200\% or more, where one is able to average the signal for several hours, at observing frequencies well above 100 GHz. The 100 MHz output produces a better result than the 10 MHz output, as might be expected. The cryocooled sapphire oscillator  has the potential to replace the hydrogen maser as the low noise frequency stable reference for millimeter wave VLBI radio astronomy.\\

\section{Acknowledgments}
This work was made possible through an Australian Research Council grant LP0883292, with support from the University of Western Australia, CSIRO, Curtin University of Technology and Poseidon Scientific Instruments. The authors would like to thank A.E.E. Rogers and S. Doeleman for useful discussions and help, M.E. Tobar for useful suggestions, and especially M. Lours for providing some necessary control circuits.

\end{document}